\documentclass[a4paper]{jpconf}
\usepackage{graphicx}
\usepackage{amsmath}
\usepackage{amssymb}
\usepackage{hyperref}
\usepackage[english]{babel}
\usepackage[utf8]{inputenc}
\usepackage[separate-uncertainty=true,
 range-units = single,
 range-phrase = -
]{siunitx}
\DeclareSIUnit\bit{bit}
\DeclareSIUnit\FLOP{FLOP}

\begin{document}
\title{Realtime processing of LOFAR data for the detection of nano-second pulses from the Moon}

\author{
T. Winchen$^1$,
A. Bonardi$^2$,
S. Buitink$^1$,
A. Corstanje$^2$,
J. E. Enriquez$^2$,
H. Falcke$^{2,5,3}$,
J. R. Hörandel$^{2,3}$,
P. Mitra$^1$,
K. Mulrey$^1$,
A. Nelles$^{2,3,7}$,
J. P. Rachen$^2$,
L. Rossetto$^2$,
P. Schellart$^{2, 8}$,
O. Scholten$^{4, 6}$,
S. Thoudam$^2$,
T.N.G. Trinh$^{4}$,
S. ter Veen$^5$  (The LOFAR Cosmic Ray KSP)
}

\address{
$^1$ Astronomy and Astrophysics Research Group, Vrije Universiteit Brussel, Pleinlaan 2, 1050 Brussels, Belgium \\
$^2$ Department of Astrophysics/IMAPP, Radboud University Nijmegen, P.O. Box 9010, 6500 GL Nijmegen, The Netherlands\\
$^3$ Nikhef, Science Park Amsterdam, 1098 XG Amsterdam, The Netherlands  \\
$^4$ KVI-CART, University Groningen, P.O. Box 72, 9700 AB Groningen, The Netherlands\\
$^5$ Netherlands Institute for Radio Astronomy (ASTRON), Postbus 2, 7990 AA Dwingeloo, The Netherlands\\
$^6$ Interuniversity Institute for High-Energy, Vrije Universiteit Brussel, Pleinlaan 2, 1050 Brussels, Belgium\\
$^7$ Now at Department of Physics and Astronomy, University of California Irvine, Irvine, CA 92697-4575, USA\\
$^8$ Now at Princeton University (USA), 4 Ivy Lane, Princeton, NJ 08544
}
	
	\ead{tobias.winchen@rwth-aachen.de}
%\author{P J Smith$^1$, T M Collins$^2$, 
%R J Jones$^{3,}$\footnote[4]{Present address:
%Department of Physics, University of Bristol, Tyndalls Park Road, 
%Bristol BS8 1TS, UK.} and Janet Williams$^3$}
%
%\address{$^1$ Mathematics Faculty, Open University, 
%Milton Keynes MK7~6AA, UK}
%\address{$^2$ Department of Mathematics, 
%Imperial College, Prince Consort Road, London SW7~2BZ, UK}
%\address{$^3$ Department of Computer Science, 
%University College London, Gower Street, London WC1E~6BT, UK}
%
%\ead{williams@ucl.ac.uk}

\begin{abstract}
The low flux of the ultra-high energy cosmic rays (UHECR) at the highest
energies provides a challenge to answer the long standing question about their
origin and nature. Even lower fluxes of neutrinos with energies above $10^{22}$ eV
are predicted in certain Grand-Unifying-Theories (GUTs) and e.g. models for
super-heavy dark matter (SHDM).  The significant increase in detector volume
required to detect these particles can be achieved by searching for the
nano-second radio pulses that are emitted when a particle interacts in Earth's
moon with current and future radio telescopes.

In this contribution we present the design  of an online analysis and trigger
pipeline for the detection of nano-second pulses with the LOFAR radio
telescope. The most important steps of the processing pipeline are digital
focusing of the antennas towards the Moon, correction of the signal for
ionospheric dispersion, and synthesis of the time-domain signal from the
polyphased-filtered signal in frequency domain.  The implementation of the
pipeline on a GPU/CPU cluster will be discussed together with the computing
performance of the prototype.
\end{abstract}

\section{Introduction}
The intensity of the flux of cosmic-rays at the highest energies (UHECR) is below one particle
per square kilometer and century. This makes answering the open questions
about their origin, acceleration, and composition a challenging task, that in particular
requires huge detector volumes~(e.g.~\cite{Kotera2011}).  The same challenge
arises in testing specific theories on super heavy dark
matter~\cite{Aloisio2006, Aloisio2015} and
grand-unification~\cite{Bhattacharjee1997} that predict an even  lower flux 
of neutrinos with energies beyond the Zetta-\si{\electronvolt} scale.

This challenge can be addressed by using Earth's moon as sensitive volume, and detect the particle interactions  
via the radio signal emitted by the Askaryan effect~\cite{Askaryan1962}.
These radio signals can possibly be recorded by radio-observatories on Earth.
However, previous searches have established only upper limits limits on the ZeV-scale neutrino flux,
and have not been sensitive enough to constrain underlying production models or
to observe UHECR~\cite{Buitink2010, Hankins1996, James2011, James2007,
Bray2014}. 

An important limiting factor for these experiments is the frequency
range of the used telescopes.  While at the used \si{\giga\hertz} frequencies
the expected pulse amplitude reaches a maximum, the pulse cannot escape the
moon for most part of its surface. The high frequencies are strongly beamed in
direction of the Cherenkov angle, while the lower frequencies are emitted in a
broader cone. As the critical angle of total internal reflection is identical to
the Cherenkov angle, only radiation which strikes the surface at an acute angle
with respect to the surface normal is emitted by the Moon. The lower
frequencies can thus reach Earth at a wider range of impact angles of the
primary particle than the higher frequencies.  As also the lunar rock is more
transparent to lower frequencies, the diminished signal amplitude is greatly
outweighed by the increase in effective detector volume. Together, this results in an
improved sensitivity of searches with telescopes operating at lower frequencies
compared to searches that used higher frequencies. The optimal frequency range
for lunar observations thus includes frequencies just above
approximately \SI{100}{\mega\hertz}~\cite{Scholten2006}.

\section{Data processing with the Low Frequency Array}
The currently largest telescope covering the optimal frequency range for
lunar detection of cosmic particles is the LOw Frequency ARray
(LOFAR)~\cite{vanHaarlem2013}, the first fully digital radio telescope.  The
antennas of LOFAR are grouped into more than 47 stations distributed throughout
the Netherlands and with additional stations in France, Germany, Poland, the
United Kingdom, and Sweden. Twenty-four of the stations are located in a dense
core with approximately \SI{2}{\kilo\meter} diameter in the Netherlands. The
additional stations are distributed with increasing distances to this core
yielding a maximum baseline of \SI{1292}{\kilo\meter}. However, due to 
technical requirements discussed below, only the core stations can be used for
lunar particle detection.

Each core station is equipped with fields of 96 low-band antennas (LBAs) with a
frequency range from~\SIrange{10}{90}{\mega\hertz} and fields of 192 high-band
antennas (HBAs) with a frequency range from~\SIrange{110}{240}{\mega\hertz}.
The signal received by the antennas is sampled in intervals of
\SI{5}{\nano\second} and copied to  a ring buffer of \SI{5}{\second} length at
antenna level before processed further. The signals of the individual
omni-directional antennas are then filtered into sub-bands by a polyphase
filter (PPF) and combined into a single beam of approximately \SI{5}{\degree}
width at \SI{120}{\mega\hertz} per station pointing towards a user-defined direction
in the sky. A selection of sub-bands is then transmitted to a
computing cluster for further processing as e.g.\ combining the station
beams to smaller `tied-array' beams and integrating the signal over longer time
spans. Each station thus assumes the role of a single dish in a classical radio
telescope.  

Detection of cosmic particles with LOFAR faces several technical challenges.
The signals of multiple stations have to be combined to maximize the sensitive
area of the antennas. As this reduces the size of the individual beam, multiple
beams have to be formed to cover the entire Moon. As each station, respectively
beam, produces data with a rate of more than \SI{6.4}{\giga\bit\per\second},
analysis of the data in real-time is required to allow only selective storage of
the data.  Signals originating from the moon are dispersed in the Earth's
atmosphere. Triggering on the amplitude of a pulse thus requires correction for the dispersion.
The pulse from a particle interaction is of only \si{\nano\second}
duration. By the PPF the time-resolution of the signal is reduced
from initially \SI{5}{\nano\second} to only about \SI{5}{\micro\second} in the
individual subbands. Here thus the inversion of the PPF is required,  which is not a
loss-less procedure. We therefore use the recovered \si{\nano\second} resolution
signal only to trigger the read-out of the Transient-Buffer-Boards containing
the original ADC traces. The offline analysis will thus not be limited by the
accuracy of the PPF inversion which may produce artefacts, but all
realtime-processing, i.e.\ beamforming, PPF inversion, and ionospheric
de-dispersion,  has to be done within \SI{5}{\second}.

\begin{figure}[tb]
	\centering
	\includegraphics[width=.6\textwidth]{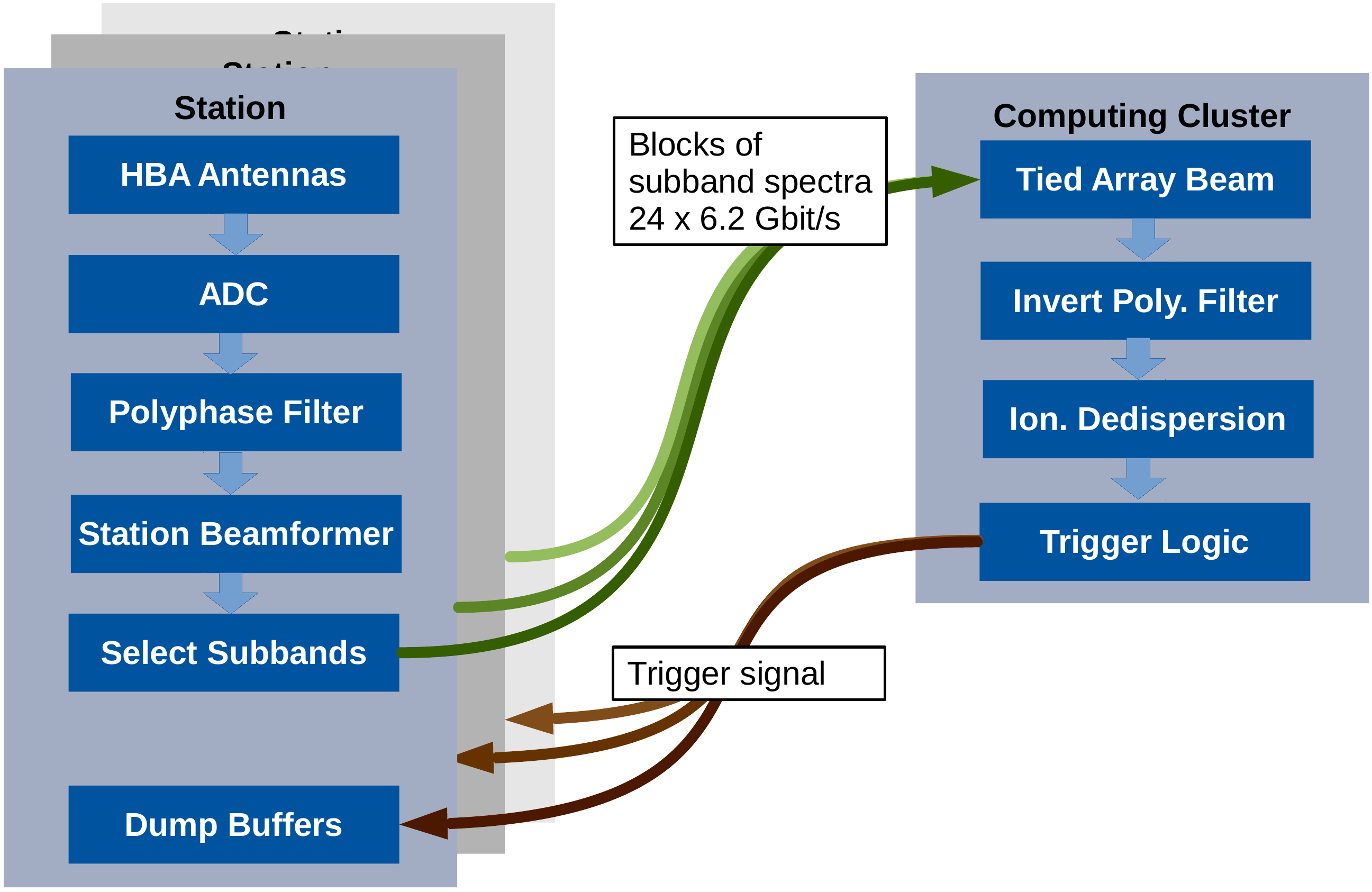}
	\caption{Overview of the online data analysis processing steps for the detection of \si{\nano\second}-pulses with LOFAR~\cite{Winchen2016a}. }
	\label{fig:ProcessingSteps}
\end{figure}
The resulting sequence of the data processing steps necessary for the lunar
detection of cosmic particles with LOFAR is sketched in
figure~\ref{fig:ProcessingSteps}. After submission to the computing cluster,
the station beams, each covering the full moon, will first be combined into
multiple smaller beam, each focussed to a small fraction of
the moon. The signal in \si{\nano\second} resolution within each beam will then be recovered
by inversion of the PPF, before the signal will be de-dispersed and eventually a trigger will be send.

\section{Computing Requirements}
The individual  processing steps are of varying computational demand as discussed in the following.  To form
a beam from the signal of $N_S$ stations, the signal of each station has to be
summed with a complex weight given by the station position and the beam
direction. We thus require $8\cdot N_S$ floating-point operations (FLOP) per sample.
With $\SI{200d6}{\per\second}$ complex samples per station, the required
computing power for beamforming is thus
$N_S\cdot\SI{1.6}{\giga\FLOP\per\second}$. 

Inversion of the polyphase filter requires an inverse Fourier transformation
(FFT) $\mathcal{F}^{-1}{\tilde{y}} = y $ of the filtered signal $\tilde{y}$ and
solving a sparse linear  system $H \hat{x} = y$ to obtain the signal $x$ with
\si{\nano\second} time resolution. In a prototype implementation~\cite{Winchen2016a} we solve the
sparse system with the iterative LSMR algorithm~\cite{Fong2011} that requires
$\mathcal{O}(100)\,\si{\giga\FLOP\per\second}$. As convergence in our tests is
achieved after approximately 25 iterations, the PPF inversion step requires
computing power of $\mathcal{O}(1000)\,\si{\giga\FLOP\per\second}$.

To correct for dispersion in the ionosphere the signal has to be multiplied in
frequency domain by a complex weight depending on the current electron content
of the ionosphere (STEC). For \SI{6}{\FLOP} per complex multiplication and $5 N
\log_2(N)$ floating point operations for a $N$-point FFT assuming the radix-2
Cooley-Tukey algorithm~\cite{Cooley1965} we thus require approximately \SI{27}{\giga\FLOP\per\second}
computing power per beam to correct for ionospheric dispersion. However, as the
current STEC is not known exactly, multiple values have to be tried to
reconstruct the original pulse amplitudes, what potentially increases
significantly the required computing power.

In total, realtime processing of a single beam thus requires several thousand
\si{\giga\FLOP\per\second} of computing power, with solving the linear
system for the PPF inversion as most costly processing step. 
%Given the
%theoretical peak performance of currently available hardware, thus either 5-10
%CPUs or 1-2 GPUs are required per beam.

\section{Available Computing and Network Resources}
 The necessary
computing power to process approximately 50 beams as required to achieve full
coverage of the moon is not available on the regular LOFAR data processing cluster
COBALT~\cite{COBALTURL}. The computations will thus be carried out on the DRAGNET
cluster~\cite{DRAGNETURL},  a CPU/GPU cluster
designed for pulsar searches consisting of 23 processing nodes. Each node is equipped with
dual Xeon Haswell-EP 2.4 GHz 8 core CPUs, 128 GB RAM, and four NVIDIA GeForce
GTX Titan X GPUs yielding a total theoretical peak performance of about
$\SI{0.5}{\peta\FLOP\per\second}$. While DRAGNET thus in principle provides
enough computational power to process the approximately 50 beams required to
cover the full moon, distributing the data over the network becomes a  
bottleneck.

\begin{figure}[b]
	\includegraphics[width=\textwidth]{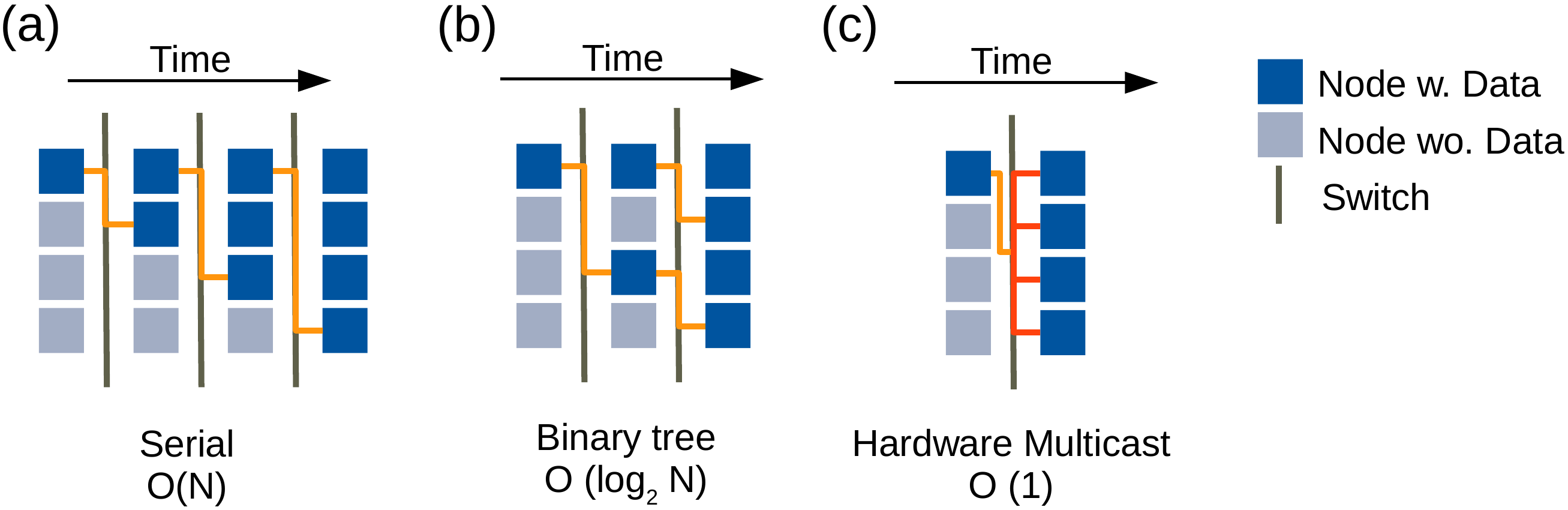}
	\caption{Different Methods of data-distribution on individual nodes connected
		to a switch. \textbf{(a)} Serial copy e.g.\ in a \texttt{for} loop.
		\textbf{(b)} Copy following a binary tree using parallel connections
		\textbf{(c)} Hardware-enabled multicast.}
	\label{fig:BroadCastTechniques}
\end{figure}
All data has to be distributed on all processing nodes on
DRAGNET\@. In a naive implementation where the data is copied in serial to $N$
nodes, e.g.\ in a \texttt{for} loop, the required bandwidth is thus increased by a
factor of $\mathcal{O}(N)$.  As switches can maintain multiple parallel
connections, copying the data to the nodes following a binary tree structure allows scaling
of the data transfer costs with $\mathcal{O}\left(\log_2N\right)$. However, the available InfiniBand switches allow also
a hardware enabled multicast~\cite{InfiniBandAssociation2015}. Here,  the data is copied on switch level and distributed
to multiple receiving  nodes in parallel, thus allow scaling in in constant time~$\mathcal{O}(1)$. These different data distribution schemes are
visualized in figure~\ref{fig:BroadCastTechniques}. 

\begin{figure}[tb]
	\centering
	\includegraphics[width=.6\textwidth]{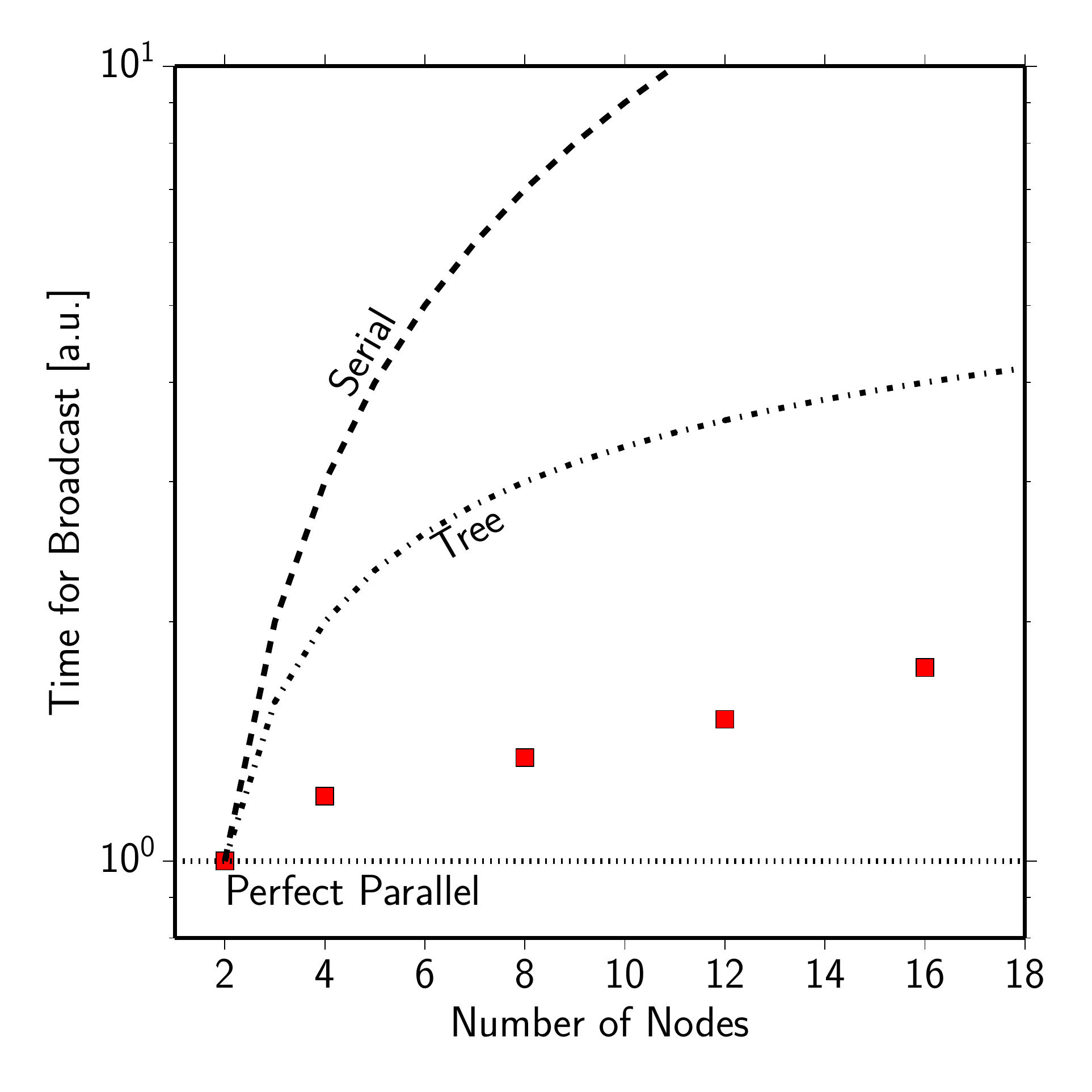}
	\caption{Broadcast performance achieved with the prototype implementation
	using OpenMPI 1.8.3 as function of the number of nodes. The total time for the copy
is normalized to the time needed for a broadcast using only two nodes, i.e.\ a single copy
operation.}
	\label{fig:BroadCastPerformance}
\end{figure}
Data distribution via hardware multicast is available for
OpenMPI~\cite{Hoefler2007}. With the default implementation in OpenMPI 1.8.3,
we achieved a speedup with a prototype implementation compared to serial or
tree level data distributions schemes as function of the number of nodes as
displayed in figure~\ref{fig:BroadCastPerformance}.  Although scaling in
constant time is not achieved, the scaling is significantly better than for
copying in a binary tree and requires less than twice the time for a broadcast
with 16 nodes than for a broadcast to a single target node. However, final
evaluation of the networking performance is still pending as the prototype was
not tested on the DRAGNET cluster but on a different cluster with similar
networking hardware to not disturb the regular operation of DRAGNET.

\begin{figure}[tb]
	\centering
	\includegraphics[width=.9\textwidth]{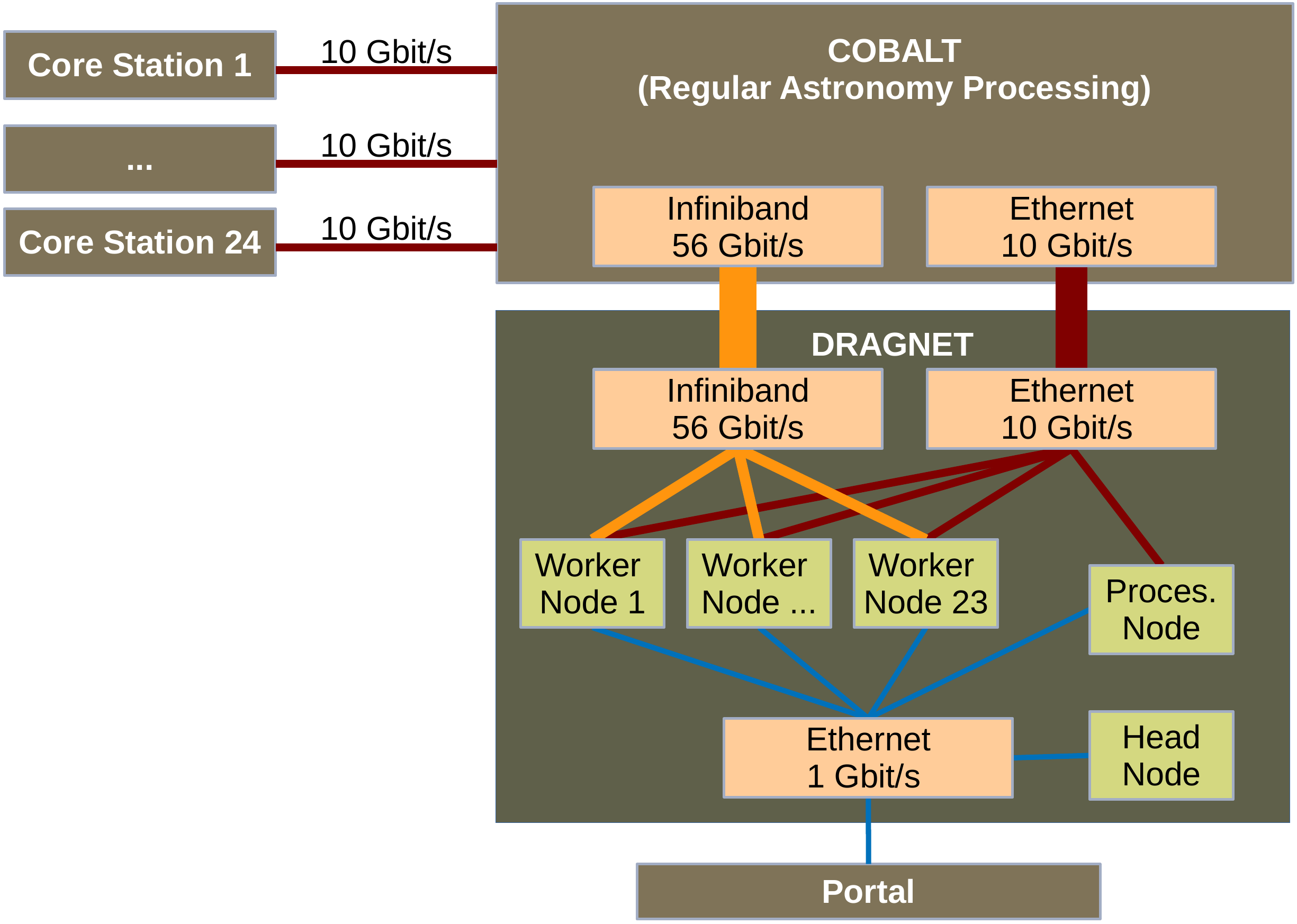}
	\caption{Schematic of the LOFAR computer network infrastructure.}
	\label{fig:LofarNetwork}
\end{figure}
The connection of DRAGNET to the LOFAR Network is sketched in
figure~\ref{fig:LofarNetwork}. Each core station is connected to the COBALT
cluster via \SI{10}{\giga\bit} Ethernet, which is sufficient to transmit the
data in the native \SI{16}{\bit} integer format yielding a data rate of
\SI{6.4}{\giga\bit\per\second} per station.  COBALT and DRAGNET are, however,
only connected via \SI{56}{\giga\bit} InfiniBand network and additionally a
\SI{10}{\giga\bit} Ethernet network. Thus even with perfect bandwidth
utilization the data of only less then ten stations can be transported to DRAGNET.
However, incorporation of data from more stations is possible by forming the beams
in two separate steps, one processed on COBALT and one processed on DRAGNET\@. In the first step,
several preliminary-beams are formed by combining the signal of only a subset of
stations on the COBALT cluster with the regular LOFAR beamformer. The data is
then transmitted to DRAGNET and the final beams are formed. The
effect of this two-stage beamforming process on the size and shape of the
final beams is currently under investigation.

\section{Prototype Implementation}
The prototype for the simulation and online software framework for lunar
observation with LOFAR is designed around the \texttt{DataChunk} as central
data container on which individual modules operate. A single \texttt{DataChunk}
either contains one polarization of the electrical field or one channel of the
digitized signal. The \texttt{DataChunk} is in a specified state, either
\texttt{TimeDomain}, \texttt{FrequencyDomain}, or \texttt{PolyphaseFiltered}.
These states define whether to interpret the stored trace as real or complex
valued and the number of samples, thus allowing the reuse of the same memory
block during processing.

Individual models operate on one or more objects of type  \texttt{DataChunk} in
one processing step and may change the state of the \texttt{DataChunk}. A
simulated PPF for example takes a \texttt{DataChunk} in the \texttt{TimeDomain}
state as input and modifies the state to \texttt{PolyphaseFiltered} after performing the
actual filter operation on the data. As another example, a module for the
simulation of the LOFAR beamformer takes as input a \texttt{DataChunk}
for the beam and adds a second \texttt{DataChunk} containing 
he signal from a different stations to the beam  
with corresponding time delay. Here, all \texttt{DataChunk}s remain in 
\texttt{PolyphaseFiltered} state.  Internally the DataChunk allocates a single array
of fixed size on the CPU and also on the GPU\@. A lazy update scheme is used to
copy the data between host and device on access to avoid unnecessary data transfer
between host and device memory while allowing code encapsulation in separated modules. 

All objects can be accessed via a python interface generated using
SWIG~\cite{Beazley1996}. The \texttt{DataChunk} is equipped with an interface
to numpy arrays.  Modules are written in C++ and CUDA for GPU processing, or
alternatively python. This allows to write efficient processing pipelines for
the simulation of individual steps or the whole experiment as well as the final
data processing pipeline as minimum python programs with a common code base.

\section{Conclusions}
To search for cosmic particles on the ZeV scale that hit Earth's moon with
earth bound radio telescopes such as LOFAR, efficient realtime processing of
the data is required. As LOFAR is optimized for astronomical observations,
significant computing resources are required to use LOFAR and the Moon as
particle detector. The necessary computing resources are, however, available
at the DRAGNET computer cluster. The now available software framework allows to
develop and test prototypes for the individual processing steps as discussed
here, and thus allows preparation of measurement campaigns to search for
particles beyond the Zette eV scale.

\section*{References}
\bibliography{astro.bib,misc.bib}

\providecommand{\newblock}{}
\begin{thebibliography}{10}
\expandafter\ifx\csname url\endcsname\relax
  \def\url#1{{\tt #1}}\fi
\expandafter\ifx\csname urlprefix\endcsname\relax\def\urlprefix{URL }\fi
\providecommand{\eprint}[2][]{\url{#2}}
% Bibliography created with iopart-num v2.0
% /biblio/bibtex/contrib/iopart-num

\bibitem{Kotera2011}
Kotera K and Olinto A~V 2011 {\em Annual Review of Astronomy and
  Astrophysics\/} {\bf 49} 119--153 (\textit{Preprint} \eprint{1101.4256})

\bibitem{Aloisio2006}
Aloisio R, Berezinsky V and Kachelriess M 2006 {\em Phys. Rev.\/} {\bf D74}
  023516 (\textit{Preprint} \eprint{astro-ph/0604311})

\bibitem{Aloisio2015}
Aloisio R, Matarrese S and Olinto A~V 2015 {\em JCAP\/} {\bf 1508} 024
  (\textit{Preprint} \eprint{1504.01319})

\bibitem{Bhattacharjee1997}
Bhattacharjee P 1997 {\em {Observing giant cosmic ray air showers from
  \textgreater 10**20-eV particles from space. Proceedings, Workshop, College
  Park, USA, November 13-15, 1997}\/} [AIP Conf. Proc.433,168(1998)]
  (\textit{Preprint} \eprint{astro-ph/9803029})

\bibitem{Askaryan1962}
Askaryan G 1962 {\em Sovjet Physics J.E.T.P\/} {\bf 14} 441

\bibitem{Buitink2010}
Buitink S {\em et~al.\/} 2010 {\em Astron. Astrophys.\/} {\bf 521} A47
  (\textit{Preprint} \eprint{1004.0274})

\bibitem{Hankins1996}
{Hankins} T~H, {Ekers} R~D and {O'Sullivan} J~D 1996 {\em Monthly Notices of
  the Royal Astronomical Society\/} {\bf 283} 1027--1030

\bibitem{James2011}
{James} C~W {\em et~al.\/} 2011 {\em Mon. Not. Roy. Astron. Soc.\/} {\bf 410}
  885--889 (\textit{Preprint} \eprint{0906.3766})

\bibitem{James2007}
James C~W {\em et~al.\/} 2007 {\em Mon. Not. Roy. Astron. Soc.\/} {\bf 379}
  1037--1041 (\textit{Preprint} \eprint{astro-ph/0702619})

\bibitem{Bray2014}
Bray J~D {\em et~al.\/} 2014 {\em Astropart. Phys.\/} {\bf 65} 22--39
  (\textit{Preprint} \eprint{1412.4418})

\bibitem{Scholten2006}
Scholten O {\em et~al.\/} 2006 {\em Astroparticle Physics\/} {\bf 26} 219 --
  229

\bibitem{vanHaarlem2013}
{van Haarlem} M~P {\em et~al.\/} 2013 {\em Astronomy and Astrophysics\/} {\bf
  556} A2 (\textit{Preprint} \eprint{1305.3550})

\bibitem{Winchen2016a}
Winchen T {\em et~al.\/} 2016 {\em Accepted for Publication\/} {\bf {in EPJ
  WoC}} (\textit{Preprint} \eprint{1609.06590})

\bibitem{Fong2011}
{Fong} D and {Saunders} M 2011 {\em SIAM Journal on Scientific Computing\/}
  {\bf 33} 2950--2971 (\textit{Preprint} \eprint{1006.0758})

\bibitem{Cooley1965}
Cooley J~W and Tukey J~W 1965 {\em Mathematics of Computation\/} {\bf 19}
  297--301

\bibitem{COBALTURL}
ASTRON 2013 Cobalt: Correlator and beamforming application platform for the
  lofar telescope (last accessed on Dec. 14, 2016)
  \urlprefix\url{https://www.astron.nl/r-d-laboratory/ska/completed-projects/cobalt/cobalt}

\bibitem{DRAGNETURL}
Hessels J 2015 Dragnet: A high-speed, wide-angle camera for catching extreme
  astrophysical events (last accessed on Dec. 12, 2016)
  \urlprefix\url{http://www.astron.nl/dragnet}

\bibitem{InfiniBandAssociation2015}
{The InfiniBand\textsuperscript{\textregistered} Trade Association} 2015 {\em
  The InfiniBand\textsuperscript{\textregistered} Trade Association
  Architecture Specification\/} vol~1

\bibitem{Hoefler2007}
Hoefler T, Siebert C and Rehm W 2007 {\em Proceedings of the 21st IEEE
  International Parallel \& Distributed Processing Symposium\/} (IEEE Computer
  Society) p 232

\bibitem{Beazley1996}
Beazley D~M 1996 {\em Proceedings of the 4th USENIX Tcl/Tk Workshop\/}
  \urlprefix\url{http://www.swig.org/}

\end{thebibliography}
%\begin{thebibliography}{9}
%\bibitem{iopartnum} IOP Publishing is to grateful Mark A Caprio, Center for Theoretical Physics, Yale University, for permission to include the {\tt iopart-num} \BibTeX package (version 2.0, December 21, 2006) with  this documentation. Updates and new releases of {\tt iopart-num} can be found on \verb"www.ctan.org" (CTAN). 
%\end{thebibliography}

\end{document}